# PACKET DROP ATTACK DETECTION TECHNIQUES IN WIRELESS AD HOC NETWORKS: A REVIEW

Kennedy Edemacu[1], Martin Euku[2]and Richard Ssekibuule[3]

[1] College of Computing and Information Sciences, Makerere University
Kampala, Uganda
[2] College of Computing and Information Sciences, Makerere University
Kampala, Uganda
[3]College of Computing and Information Sciences, Makerere University
Kampala, Uganda

## ABSTRACT

*Wireless ad hoc networks have gained lots of attention due to their ease and low cost of deployment. This has made ad hoc networks of great importance in numerous military and civilian applications. But, the lack of centralized management of these networks makes them vulnerable to a number of security attacks. One of the attacks is packet drop attack, where a compromised node drops packets maliciously. Several techniques have been proposed to detect the packet drop attack in wireless ad hoc networks. Therefore, in this paper we review some of the packet drop attack detection techniques and comparatively analyze them basing on; their ability to detect the attack under different attack strategies (partial and or cooperate attacks), environments and the computational and communication overheads caused in the process of detection.*

## 1. INTRODUCTION

Wireless ad hoc networks are a group of computing devices equipped with radio transceivers and interconnected wirelessly through radio frequency without a fixed infrastructure or centralized control [13]. The most common examples of wireless ad hoc networks are wireless sensor networks [1] [22] and mobile ad hoc networks [2].

In wireless ad hoc networks, nodes communicate with each other using multi hop wireless links. Data to out of range nodes can be routed through intermediate nodes. That is nodes in wireless ad hoc networks can act as both hosts and routers. There are numerous application areas in which wireless ad hoc networks can be used ranging from military operations and emergency disaster relief to community networking and interaction among meeting attendees or students during lectures [21].

One basic assumption according to Djahel *et al* [3] in design of routing protocols in wireless ad hoc networks is that, every node is honest and cooperative. This introduces a vulnerability that can be exploited for launching attacks. An example of such attack is the malicious packet dropping.

A number of techniques have been proposed in literature to mitigate the packet dropping problem in wireless ad hoc networks. In this paper, we review some of the malicious packet dropping detection techniques proposed in literature and comparatively analyze them. The rest of the paper is organized as follows; section 2 discusses packet dropping in wireless ad hoc networks, in section 3, we discuss some of the packet drop attack detection techniques proposed, section 4 presents a malicious node identification tool, section 5 compares these techniques under different scenarios and section 6 concludes the paper.

# 2. PACKET DROPPING IN WIRELESS AD HOC NETWORKS

Like in any other network, packet loss is expected in ad hoc networks at least to an acceptable percentage. Not all packets lost should be viewed as malicious. In this section, we discuss some of the packet loss scenarios in wireless ad hoc networks.

## 2.1 Legitimate Packet Dropping

Packet dropping can be experienced in wireless ad hoc networks where no compromised nodes are present. This packet loss is mainly associated with the following events;

### I. Network Congestion

Network congestion in wireless ad hoc networks is something unavoidable. These networks are mainly scalable due to in and out movements of nodes. As a result, congestion is more likely to happen which can lead to loss of packets.

### II. Channel Conditions

In wireless networking the channel condition cannot be neglected since it changes drastically. Free path loss, interference, presence of noise on the channel and fading of the transmitted wireless signals are among the channel conditions that can lead to packet loss or bit errors in the transmitted signal. In the presence of these factors, some packets can get dropped.

### III. Resource Constraints

Nodes in wireless ad hoc networks have limited energy resource [23] [24] [30]. Intermediate nodes in these networks may behave selfishly and fail to forward the received packets in order to conserve their limited resources battery power. These packets in turn get dropped.

## 2.2 Malicious Packet Dropping

Mostly, the first step in launching a packet dropping attack is for a malicious node to get involved during route formation. This is better done by exploiting the vulnerabilities of the underlying well known routing protocols used in wireless ad hoc networks which are designed basing on the assumption of trustworthiness between nodes in a network.

Once in the route, the malicious node can do anything including maliciously dropping packets. This Packet dropping at a malicious intermediate node can lead to suspension of communication or generation of wrong information between the source and destination which is an undesirable situation. For better understanding, following are the illustrations of malicious packet dropping scenarios in wireless ad hoc networks under the two commonly used routing protocols AODV (Ad hoc On Demand Distance Vector) and OLSR (Optimized Link State Routing).

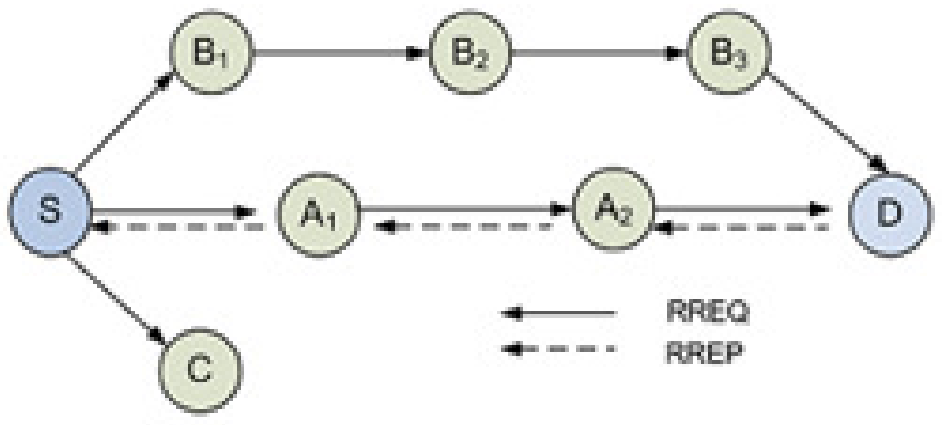

Figure 1. Route Discovery in AODV [4]

## A. Packet Dropping in AODV

The route discovery process between source (S) and destination (D) under AODV routing protocol is as illustrated in Figure 1.The source broadcasts a RREQ (Route Request) message with unique identifier to all its one hop neighbors. Each receiver rebroadcasts this message to its one hop neighbors until it reaches the destination.

The destination on receiving the message updates the sequence number of the source and sends a RREP (Route Reply) message back to its neighbor which relayed the RREQ. On the other hand, an intermediate node that has a route to the destination with destination sequence number equal to the one in RREQ can send back a RREP packet to the source node without relaying to the destination.

For a node to launch packet dropping attack, it must be involved in at least one routing paths in the network. This is illustrated in Figure 2, C is a malicious node intending to drop packets from S to D. To discover a path from S to D, S first broadcasts RREQ packet to its neighbors. Each neighboring node continues to rebroadcast this message as explained earlier until it reaches D.

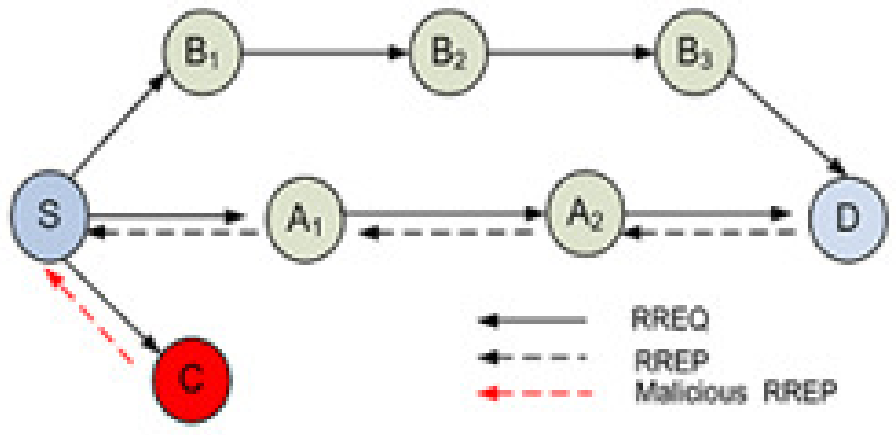

Figure 2. Packet Dropping Attack in AODV [4]

The malicious node C disobeys this rule and lies to S claiming it has the shortest path to D and sends a RREP packet to S. As a result, S assumes that the shortest route to D is through C and starts to send data packets to D through C which are in turn dropped.

## B. Packet Dropping in OLSR

OLSR uses Multipoint Relays (MPRs) which are set of neighboring nodes that are responsible for spreading the local link state information to the whole network for optimization. The link state is broadcasted periodically through Topology Control (TP) messages. Each node in OLSR selects its MPR set from its one hop neighbors such that it can easily reach all its two hop neighbors with minimum number of retransmissions.

Selection of the MPR depends on the number of two hop neighbors reachable through the candidate node and its “Willingness value obtained from “Hello” message which indicates the readiness of a node to forward packets of its neighbors.

Through periodic exchange of link state, each node senses its neighbors and disseminates the network topology. Each node constructs a partial topology graph of the network from broadcasted TC messages which allows it to establish routes to non-neighboring nodes.

For a packet dropping attack, a malicious node may send a TC message claiming to be a MPR of nodes although it may not. As the network depends on the MPRs for routing services, the malicious node may decide to drop packets passing through it.

## 3. PACKET DROPPING ATTACK DETECTION TECHNIQUES

A number of malicious packet dropping detection techniques have been proposed in literature. In this section we discuss some of them;

### 3.1 Watch Dog Technique

The watch dog technique has been the most well know node misbehavior detection in ad hoc networks. In this technique, every node acts as a watchdog agent monitoring packet transmissions to neighboring nodes [17]. The watchdog agents save a copy of packets in their watchdog monitoring buffers before their transmission to the next node. This serves to monitor packet relay from a neighboring node to the next node.

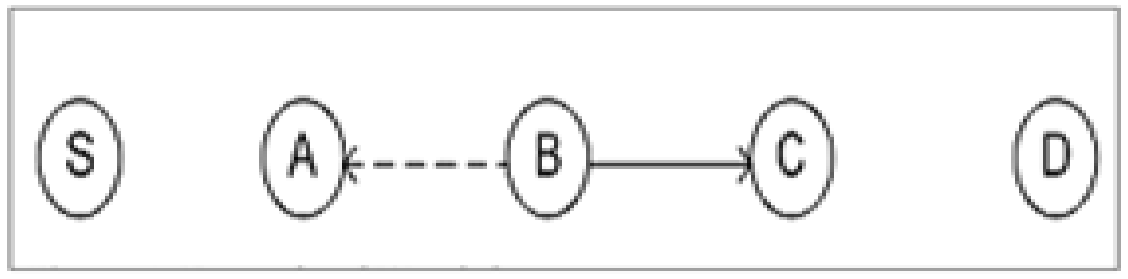

Figure 3. An illustration of Watch Dog [17]

Figure 3 shows an example of watchdog. "S" is the source node and "D" is the destination node. The other nodes are intermediate nodes in the route between 'S' and 'D'. Before 'A' forwards a packet received from 'S', it saves the packet in its watchdog monitoring buffer. After forwarding the packet to 'B', 'A' monitors whether the packet has been forwarded to 'C'. This is because 'A' is expected to receive a copy of the packet forwarded to 'C' since it's within 'B's transmission range. 'A' then compares the received packet with the one saved in its watchdog monitoring buffer. If 'A' fails to receive a copy of the packet from 'B' within certain duration, it reduces the confidence level of 'B' by 0.05. When this happens in recurring manner, the confidence level is set to zero and 'A' decides that 'B' is a malicious node and sends an alarm so as to change the route through 'B'. Meanwhile if 'B' forwards the packet within the time duration, 'A' rewards 'B' by increasing its confidence level by 0.01.

Forootaninia *et al* [10], proposed an improved version of watch dog technique (I-Watch dog). In I-Watch dog technique, the cluster head is assumed to be the watchdog as illustrated in Figure 4.

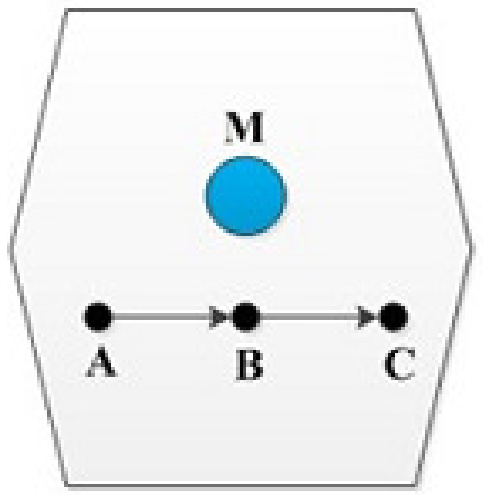

Figure 4. Proposed Improved-Watchdog (I-Watchdog) [10]

Incase node **A** wants to communicate with node **C**, node **M** which is the cluster head acts as the watchdog. It's assumed that, **M** has a buffer that accommodates all the sent packets within its range. Node **B** is the intermediate node between **A** and **C.** **M** monitors the packet forwarding character of **B** after receiving **A**'s message which is stored in its buffer and compares the received copy from **B** with the buffered message. If the messages are similar, the buffered message is dropped. Otherwise, it is considered that, **B** has not sent the message or replaced with another. This work though improves the life span and power consumption of sensor nodes, has not addressed the packet dropping detection inefficiencies of watch dog technique.

Under circumstances of ambiguous collision problems, False misbehavior, receiver collision problems insufficient transmission power, Cooperated misbehavior and partial packet dropping [25] [27], watch dog technique fails. Since watch dog relies on confidence values, a malicious node can partially drop packets so that its confidence value does not come to zero. This technique is only suitable in environments in which the watch dog agent has knowledge of its two hop neighbors [26]. As a result, it is only more effective in source routing protocols like Dynamic Source Routing (DSR).

## 3.2 Side Channel Monitoring (SCM)

In SCM a sub-set of neighbors for each node in a route between source and destination are selected to observe and monitor their message forwarding behaviors [18]. Alarm channel (Primary channel and Side channel) is generated to inform the source about the misbehaving node; The Primary channel (PC) is formed by nodes in the route and Side channel (SC) is formed by sub-set of monitoring neighbors.

As an example, consider an established data communication path R = $a_0 \ldots a_{k+1}$ with $k \geq 0$ and $a_0$ , $a_{k+1}$ being source and destination respectively. It's assumed that $a_1, \ldots, a_k$ are normal at the route establishment time and that $a_0$ and $a_{k+1}$ are always normal during their communication time.
Packets transmitted along R can be classified as data packets or alarm packets. Data packets flow from $a_0$ to $a_{k+1}$ meanwhile alarm packets flow in the opposite direction. The alarm packets are normally generated by intermediate nodes and transmitted to the source through the SC.
In (Figure 5a), for the two successive nodes $a_{i-1}$ and $a_i$ between the source and destination, a set $G_i$ of nodes are selected as observers. The selected nodes in $G_i$ are responsible for monitoring whether $a_i$ forwards $a_{i-1}$'s data packets towards $a_{k+1}$ and generate an alarm when it fails to forward within a given time window. In Figure 5b, the logical view of observer nodes is presented. $G_i$ [1 2 3 4 5] monitors $a_{i-1}$ and $G_{i+1}$ [4 5 6 7 8 9] monitors $a_i$.

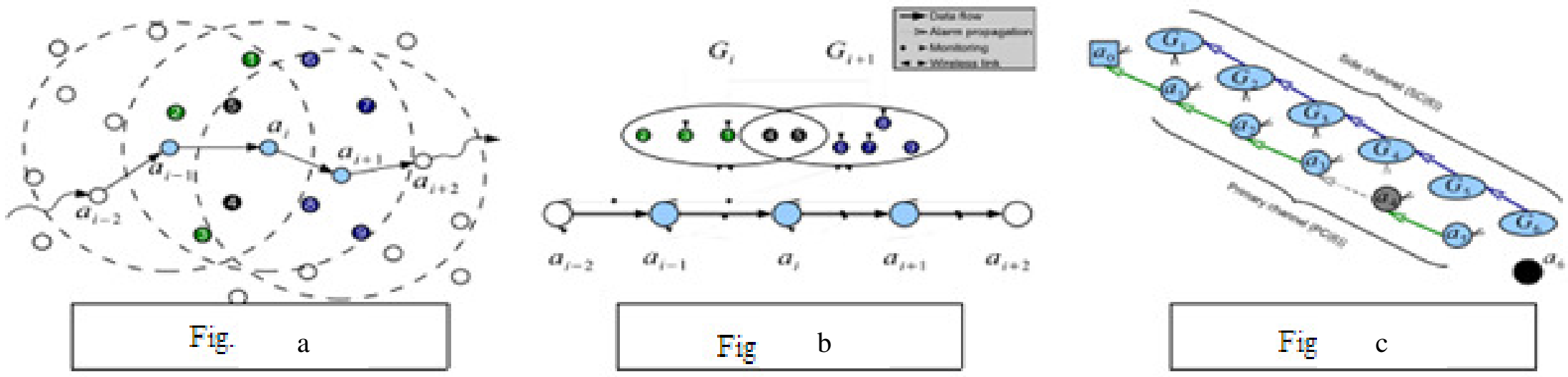

Figure 5. An illustration of Side Channel monitoring Technique [18]

For a particular node under observation, the set of monitoring nodes keep observing its data forwarding behavior. If it fails to forward a data packet within a given time period, the set generates an alarm that is sent to the source through the SC created between different sets of monitoring nodes as illustrated in Figure 5c. The watch dog nodes in the path also construct a PC for forwarding such kind of misbehavior to the to the source.

The source accumulates this information about a particular node and once it has exceeded a threshold value, it generates a warning message that is forwarded to all the nodes in the network and reconstructs a different path that excludes the misbehaving node.

Since, the primary channel operates the same way as the watch dog and all the neighboring nodes have the ability to do side channel monitoring, in mobile environments, SCM detects packet drop attack. SCM on the other side generates lots of network traffic from the side channels and primary channel causing communication overhead.

### 3.3 Monitoring Agent Technique

[19], proposed the monitoring agent technique. The technique is based on capturing packets sent by neighboring nodes within a transmission range. All the nodes in a network collect information about their one hop neighbors within a certain period of time. The collected information include; the total number of packets transmitted from a particular node (*WLi*), the average number of transmitted packets from all its one hop neighbors (*AWL*), the packet drop rate of a particular one hop neighbor (*DRi*), and the average packet dropping rate by all its one hop neighbors (*ADR* ) which are used for identifying a malicious node.
Figure 6 is an illustration of the concept and S is the monitoring agent. S uses information collected from its neighbors to determine whether there are legitimate or malicious nodes. Assuming the information S collected during a certain period from its neighbors is as in table 1.

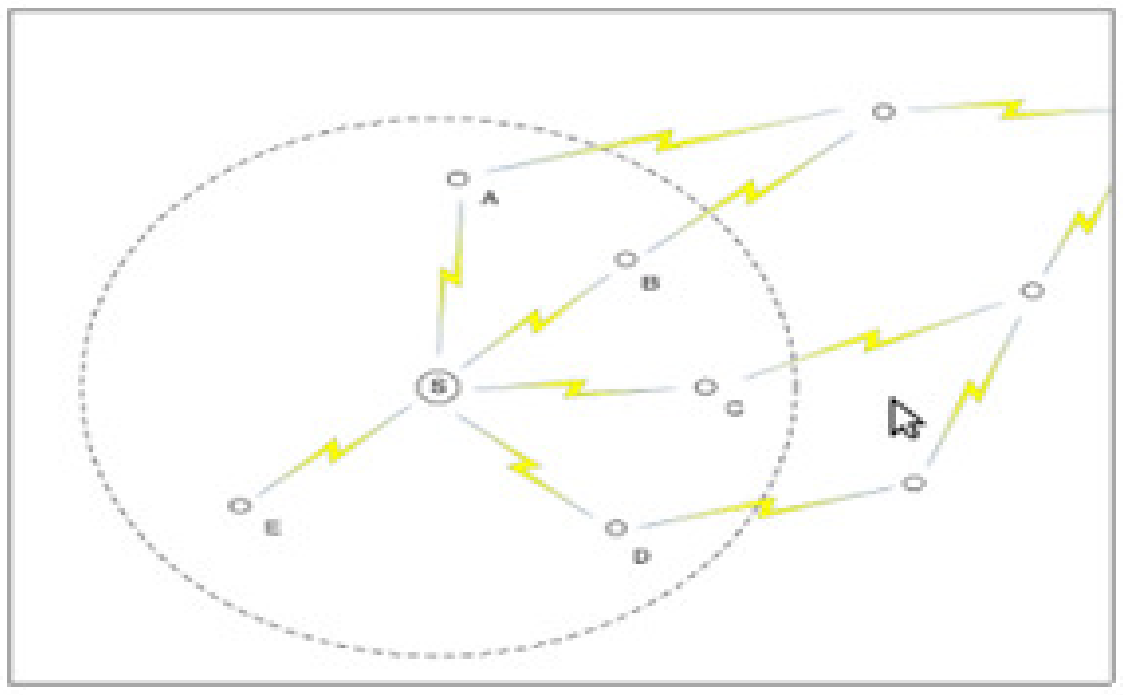

Figure 6. An illustration of monitoring Agent [19]

| | Node A | Node B | Node C | Node D | Node E |
|---|---|---|---|---|---|
| $WL_i$ | 100 | 192 | 281 | 296 | 351 |
| $DR_i$ | 80% | 26% | 19% | 40% | 37% |

Table 1. Measured sample values [19]

From table 1, node A is assumed to be a malicious node due to its high packet drop rate.
Another parameter *DEVi* (*ADR-DRi*) helps in determining a suspicious node. *DEVi* is calculated for each node before sending a packet from the monitoring node to each neighboring node. The nodes are then sort in a descending order based on *DEVi* to classify the suspicious and normal nodes as illustrated in the table 2.

| | Node C | Node B | Node E | Node D | Node A |
|---|---|---|---|---|---|
| *DEVi* | 21.4% | 14.4% | 3.4% | 0.4% | -39.6% |

Table 2. *DEVi* sorted in descending order [19]

Also a parameter *Dn* (Average packet drop rate (*ADR*) /2) is calculated. If a node has *DEVi* more than *Dn*, it's assumed to be normal otherwise it's considered malicious.

To reduce false negatives (whether a node dropped packets maliciously or due to traffic problem), *AWL* (average of packet transmission by neighboring nodes) is calculated. In table 1, the *AWL* during a certain period of time is 224. The total number of packets broadcasted by a malicious node A during that period was 100 and this was lower than the average number of packets, so it had a high packet drop rate. Thus, monitoring node S determined that node A was maliciously dropping packets and an alarm message was sent to the entire network so as to inform the source.

## 3.4 Sequence Number Model

In [20], another technique "Sequence number model technique" for detecting packet dropping attack was proposed. Consider Fig 6, source (S) is out of communication range with destination (R) and uses intermediate node 'X' to reach 'R'. When 'R' determines that the rate of packet loss using TCP sequence number through 'X' has exceeded a certain threshold value, 'R' becomes suspicious and starts the investigation.

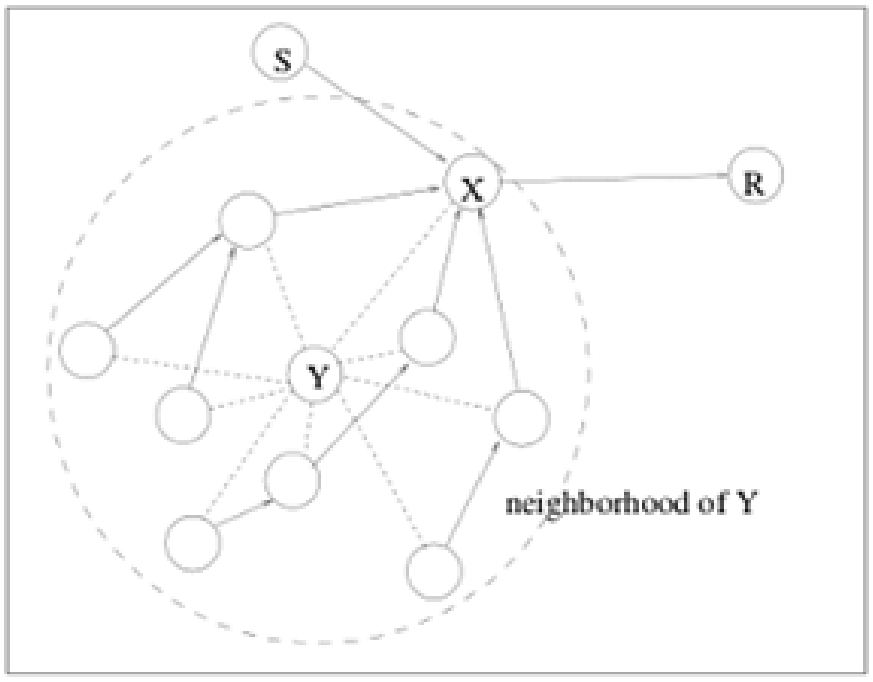

Figure 7. An illustration of Sequence number model [20]

'R' first sends 'X' a message requesting it to broadcast a special message to all its neighbors. The special message requests all X's neighbors to send 'R' a secure message communicating the traffic amount they forward to 'X'. Once 'R' sends the special message to 'X', it starts to monitor 'X' to determine whether it will send the message to its neighbors. To prevent an intruder from making X's neighbors sending 'R' false message in response to its query, an authentication system based on private symmetric session key is used. Once 'R' gathers all the information from X's neighbors, it can determine whether the packet loss is due to congestion or malicious.

If another node other than 'X', say 'Y' is the one compromising the network, it may decide to send false information to 'R'. To overcome this, each node is asked to monitor the total traffic sent by each of its neighbors. This helps 'X' to identify the true intruder 'Y' and informs 'R' about it.

### 3.5 PathRater

PathRater is run by every node in the network [31] [26]. A node maintains ratings for every other node it knows in the network basing on the knowledge of misbehaving nodes and link reliability data in order to choose the most suitable path. A path metric is calculated by averaging the ratings for nodes in the path. In case there is more than one path to the destination, the path with the highest metric is chosen. A pathrater node assigns a neutral rating of 0.5 to nodes known to it. It normally assigns itself a rating of 1. The ratings are updated in intervals of 200ms. The ratings for nodes in active path are increased by 0.01 and the maximum rating a node can attain is 0.8. A node's rating is decreased by 0.05 when a link break is detected and the node becomes unreachable.

A negative path metric value indicates presence of misbehaving nodes in the path. Due to faults or false accusations, a node may be marked as a misbehaving node. It is generally better not to permanently mark it as misbehaving node. Therefore, the marked misbehaving nodes' rating should be increased slowly or set back to 0.0 after a long time period.

When a pathrater detects a misbehaving node in a path it is using and fails to get an alternative path free of misbehaving nodes, it sends out a route request message called Send Route Request (SRR) [31]. This way, a new metric can be constructed from which a new path can be determined.

### 3.6 TwoAck

In this technique, packets sent by a node are expected to be received by nodes which are two hops away in the path [32]. Nodes in a path are expected to send acknowledgement packets called TWOACK packets two hops backwards. If a node fails to receive TWOACK packet after sending or forwarding packets, the next node's link is considered to be misbehaving and will be eliminated in the next routing.

In order to reduce the overhead due to these acknowledgement messages, a scheme called selective-TWOACK (S-TWOACK) which selectively acknowledges packets was proposed in [33]. In this scheme, an acknowledgement is sent after receiving certain number of data packets.

## 4. MALICIOUS NODE IDENTIFICATION

Not all the packet dropping detection techniques can identify malicious nodes. Some simply detect the packet dropping misbehavior without identifying the malicious node. Once malicious packet dropping behavior is detected, the responsible malicious node needs to be identified. There is one malicious node identification tool (traceroute tool) being used in wired networks which can be adopted for wireless ad hoc networks [28].

### 4.1 Traceroute

Traceroute [29] is used to determine the route to a destination node. It sends probing UDP messages with increasing Time-to-Live (TTL) values towards the destination. The first probe is assigned a TTL value of 1. This enables the message to travel only the first hop in the path to destination. Once the TTL elapses, the packet gets dropped and an ICMP time exceeded message is sent back to the probing node. The probing node then keeps on increasing the TTL value for the successive probing UDP messages until it reaches all the intermediate nodes between the source and the destination. Every time the UDP packet gets dropped, an ICMP port unreachable message is sent back to the source as illustrated in Figure 8.

In Figure 8, node 1 wants to reach node 9. It sends probing UDP messages with increasing TTL values until it reaches the destination node 9. Every time a TTL elapses in an intermediate node, the node sends back an ICMP time exceeded (TE) message to node 1 with its address. This way, node 1 is able to know the route to node 9 by combing the received addresses from the respective ICMP messages.

In case a malicious packet dropping node is in the route between the source and the destination, the sent UDP messages get dropped and fail to reach the destination. Since, for every dropped packet an ICMP message containing the address of the dropping node is sent back to the source, it is possible to isolate and identify the malicious node.

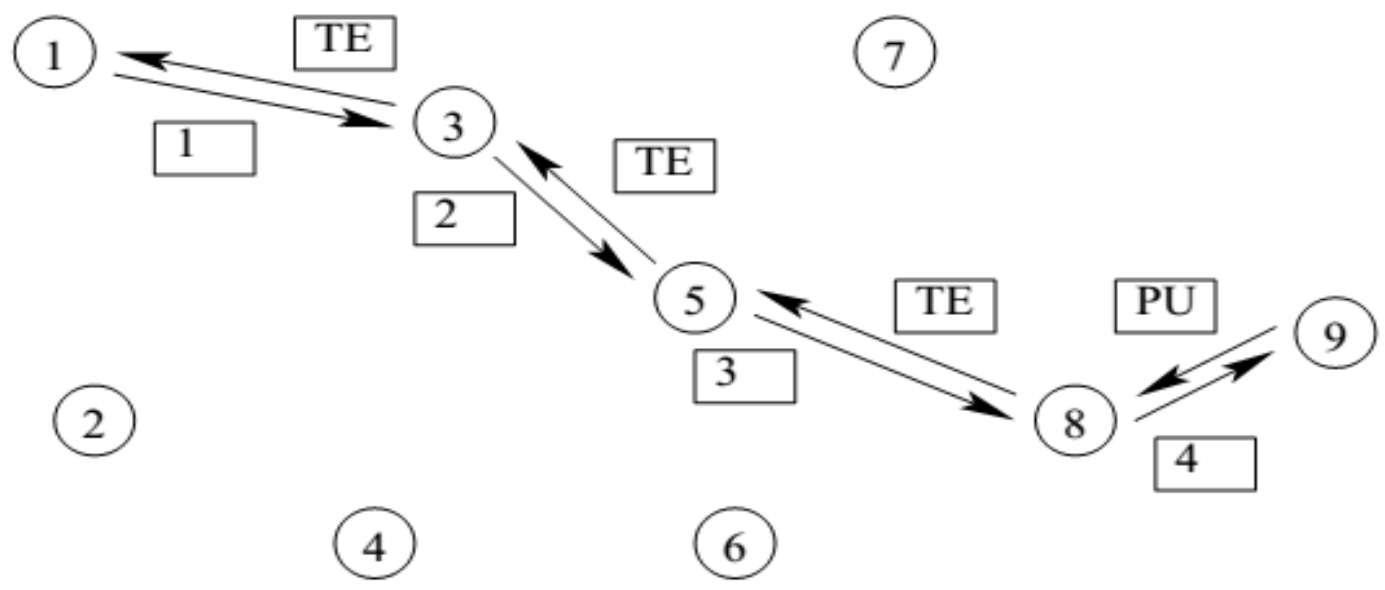

Figure 8. An Illustration of traceroute [28].

## 5. COMPARISON OF THE PACKET DROP ATTACK DETECTION TECHNIQUES

In this section, we compare the discussed packet dropping attack detection techniques in terms of; the attack kinds they can detect (partial and cooperate attacks), the potential to detect attacks in mobile environments, the computational overhead and the communication overhead caused.
Some malicious packet dropping detection techniques in wireless ad hoc networks consider a node malicious if it drops packets above a certain threshold value, the partial attack exploits this principle in that the malicious node if aware of the threshold will drop packets to a value just below the threshold. On the other hand, cooperate attack is one where two or more coordinated malicious nodes are used to launch an attack. The malicious nodes may be on the same or different routes but usually are synchronized.

The comparison for the different packet dropping attack detections are illustrated in table 3.

| **Technique** | **Computational Overhead** | **Communication Overhead** | **Mobility Environment** | **Partial Attack** | **Cooperate Attack** |
|---|---|---|---|---|---|
| Watch Dog | Low | Low | Yes | No | No |
| SCM | Average | Average | Yes | Yes | Yes |
| Monitoring Agent | High | Average | No | Yes | Yes |
| Sequence Number Model | Average | High | No | Yes | No |
| PathRater | Average | Low | Yes | No | Yes |
| TwoAck | Low | High | Yes | Yes | Yes |

Table3. Comparison of the different Packet dropping Attack Detection techniques

## 6. CONCLUSION

Wireless ad hoc networks are widely used in military and civilian applications. Security which is a critical factor is a concern in these kind of networks due to lack of centralized control. This results into launching of different attacks including the packet dropping attack in these kind of networks. In this work, we reviewed the malicious packet dropping in wireless ad hoc networks considering the two common routing protocols AODV and OLSR as example cases. We comparatively analyzed some of the malicious packet dropping detection techniques proposed to assess their effectiveness and limitations. The Watch Dog though produces less computational and communication overhead is susceptible under partial and cooperate attacks. The SCM produces average computational and communication overheads as it detects different kinds of attacks in mobile environments. The Monitoring Agent produces high overheads in detecting different attack kinds and is susceptible in mobile environments. And the sequence number technique as well produces high overheads and is susceptible in mobile environments and does not detect all attack kinds.

Depending on the anticipated attack strategy, network environment (mobile or stationary) and the processing power of the nodes to be used in a given wireless ad hoc network the choice of a malicious packet dropping technique can be guided by this work. As a future work, designing a malicious packet dropping detection technique that effectively detects the packet dropping attack in any environment while keeping the generated overheads minimal will be our focus.

## Authors

**Kennedy Edemacu** received Bsc in Computer Science from Gulu university in 2011. He has received Msc in Data communication and Software Engineering from Makerere University. He is currently working as a teaching staff in Muni University. His recent research interest s include; Machine-to-Machine communication, Security in Wireless Ad hoc Networks, Resource Allocation and Image Processing.

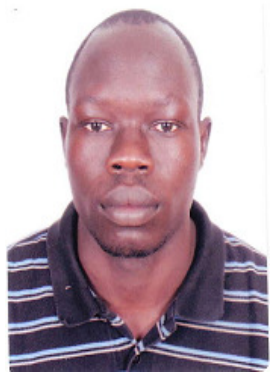

**Martin Euku** received Bsc in Computing and Information technology from Kyambogo University in 2011. He is currently pursuing Msc in Data Communication and Software Engineering from Makerere University. His research interests include; Security in Ad hoc networks and Long Distance Wi-fi

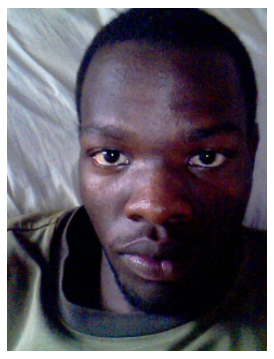

**Richard Ssekibuule** is pursuing Phd from Makerere university. He is currently teachin g in the college of Computing and information sciences Makerere University. He has several publications in International Journals and Conferences. His current research interests include; Cryptography, Machine Learning and Auction Mechanisms.

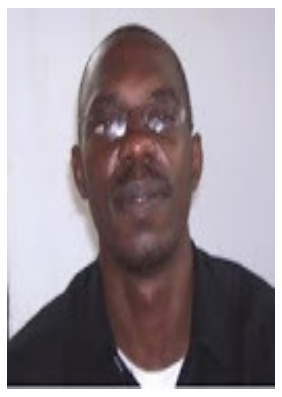